\documentstyle[12pt,prd,aps,psfig,floats,preprint]{revtex}

\begin{document}
\title{Modelling the propagation of an ultrasonic ray through
a heterogeneous medium}

\author{Carlos R. Fadragas\thanks{fadragas@mfc.uclv.edu.cu},
Manuel Rodriguez\thanks{mrg@mfc.uclv.edu.cu}, Rolando
Bonal\thanks{rolandob@mfc.uclv.edu.cu}}

\address{Department of Physics, School of Mathematics,
Physics and Computer Science,\\ Central University of Las Villas,
Santa Clara, Cuba}

\date{\today}
\maketitle \draft

\begin{abstract}
The estimation of mechanical properties of concrete is one of the
most important applications of the ultrasonic method. This
technique often involves the measuring of the ultrasonic pulse
propagation velocity. However, there exist some factors
influencing the effective ultrasonic pulse propagation velocity in
concrete. This paper is devoted to a discussion of the behavior of
the effective ultrasonic pulse propagation velocity as a function
of the free water contained in concrete. The travelling of an
ultrasonic ray through a heterogeneous medium composed by cement
and aggregates(solid phase), water(liquid phase) and air(gaseous
phase) was empirically modelled. A relationship between the
transition time of the ultrasonic ray in the testing piece and the
free water quantity was considered. We assumed an empirical
intermediate relationship between the ultrasonic path length in
porous space in concrete filled with water and the free water
quantity in the testing piece. From experimental data points we
evaluated the parameters of the model and obtained a fitting
function for describing the dependence between the effective
ultrasonic pulse propagation velocity and the free water quantity
in concrete. We also analyzed the error propagation through the
algorithm used. It allowed to decide under which conditions the
linear approach between the effective ultrasonic pulse propagation
velocity and the free water quantity in concrete can be applied.
\end{abstract}

\newpage
\section{Introduction}
The estimation of mechanical properties of concrete, in
particular, that of compressive strength, is one of the most
important applications of the ultrasonic method
\cite{Angel,DiMaio,Ferreira,Ibrahim,Manuel,Rhoon}. However, the
effective application of this method is subject to the influence
of several factors \cite{Manuel,Katheen,Sandor,ASTM,British}. One
of these factors is the water which penetrates in concrete through
the pores \cite{Hall,HallC,Kelham,Carter}. Under the influence of
this factor, the compressive strength and the effective ultrasonic
pulse propagation velocity differently behave as the free water
content changes. In fact, the latter increases as the free water
content in the specimen increases. On the other hand, the
compressive strength practically behaves constant when the free
water content is modified\cite{DiMaio,Rodriguez}.\\

Under these circumstances, some considerations will be made. The
free water in concrete is one of the most important building
problems widely considered today. The concrete strength is
generally considered its most important property among other
characteristics for establishing its quality. This property of the
concrete may be influenced by its free water content. Several
authors consider that the durability of a building structure is
determined by the rate at which water and chemical components that
it contains infiltrate and move through the porous building
materials\cite{Kuntz}.\\

On the other hand, the ultrasonic transmission method is often
used for estimating, among other properties, the concrete
strength. The application of the ultrasonic transmission technique
to measuring the ultrasound propagation velocity through concrete
allows to estimate the value of concrete strength\cite{Manuel}.
The relationship between these two quantities is often established
when the concrete in service is dried. When free water infiltrates
and moves through the pores the measuring of the effective
ultrasonic pulse propagation velocity can not be directly used for
estimating the value of the concrete strength. It is still an open
problem. However, the application of the ultrasonic transmission
method for determining the ultrasound pulse velocity can allow to
estimate the free water content in the porous structure but it can
not allow to decide about the damage of the building. The free
water content affects the ultrasonic pulse propagation velocity
but this measure says nothing about the current strength of
concrete. If the dried state of concrete can be obtained back
again, then the measuring of the ultrasonic pulse propagation
velocity made in that condition could indicate a possible change
in the value of concrete strength. This problem has a high
attracting social attention in a tropical country where several
types of meteorological events often produce a high level of
moisture in building during several days. M.
Rodriguez\cite{Manuel}has studied the influence of moisture in
building in tropical condition finding that this phenomenon
produces a significatively high
damage in building structure.\\

 M. Kuntz and P. Lavall\'ee\cite{Kuntz} have studied the infiltration
 of liquid and the propagation of the moisture front in non-saturated porous
media using an one-dimensional diffusion equation for describing
the phenomenon. They considered a generalized scaling law of the
type $x t^{-\alpha}$ with $\alpha=1/(n+1)$ and being $n \in \Re$.
The water transfer in partially saturated materials was assumed to
follow a general nonlinear diffusion equation being water content
the variable to determine. They found the values $\alpha=0.58$ in
brick and $\alpha=0.61$ in the limestone diverging both from the
value $\alpha=0.5$. The results mean that the volume of absorbed
water in this materials may be underestimated, having a particular
importance for evaluating the effect of water content in building
structures. This result expresses that there exists a great
theoretical interest about the complexity of this problem along
with the practical approach.\\

E. Ohdaira and N. Masuzawa\cite{Masuzawa} have examined the
possibility of the NDE of concrete from a knowledge of the
behavior of the ultrasound propagation velocity with water content
in concrete. They measured the ultrasound velocity and the
frequency component on ultrasonic propagation as a function of
water content in concrete. They made testing pieces of concrete
being 10 cm in diameter and 20 cm in length. These testing pieces
were embedded in water for 50 days. After this, the water in each
test piece was gradually extracted and the ultrasound propagation
velocity was conveniently measured for having 15 points in the
relationship between water content and the ultrasound propagation
velocity. They found that the ultrasound propagation velocity
decreases approximately linearly in proportion to the decrease in
water content. Similarly the same occurs for the frequency of the
maximum transmission.\\

M. Rodriguez and R. Bonal\cite{Bonal} made a comment on the paper
referred above. They considered that the relationship between the
ultrasound propagation velocity and water content can better be
described by using a statistical model. They fitted the
experimental data points obtained by Ohdaira and
Masuzawa\cite{Masuzawa} by applying this method and obtaining an
empirical fitting equation. They concluded that the ultrasound
propagation velocity exponentially depends on the free water
quantity in concrete.\\

In this work an empirical nonlinear model for the relationship
between the effective ultrasonic pulse propagation velocity $v$
and the quantity of free water in concrete ($m_{w}$) was
considered by us to better describe that behavior. The
experimental data given in \cite{Masuzawa} were considered again.
It was also discussed under which conditions the linear approach
can be used.

\section{Modelling the travelling of an ultrasonic ray through
a heterogeneous medium}

We consider the experimental disposition scheme for applying the
transmission technique in ultrasonic method to determine the
ultrasonic pulse propagation velocity. An ultrasonic ray
travelling through a concrete testing piece containing free water
can be modelled by considering an equivalent scheme consisting of
a heterogeneous medium composed by three phases of the matter:
gaseous phase (air), liquid phase (free water), and solid phase
(fine and coarse aggregate, cement). The total transmission time
$t$ of the ultrasonic ray in the acoustical system may be
decomposed as

\begin{eqnarray}
t=t_a+t_w+t_s+\delta
\end{eqnarray}

where we have separately considered the time interval spent by the
ultrasonic ray for travelling through each phase in the
heterogeneous medium: $t_a$ corresponds to the air, $t_w$
corresponds to the free water, and $t_s$ corresponds to the solid
phase of concrete. Finally, $\delta$ represents the time interval
spent by the ultrasonic ray for travelling outside the testing
piece, and it will be considered a constant value during the
measuring procedure on a given testing piece. Its numerical value
in all these experiments was taken to be $\delta=0.6 \mu
s$\cite{Masuzawa}. The equation (1) may be written as

\begin{eqnarray}
t-\delta=\frac{h_a}{v_a}+\frac{h_w}{v_w}+\frac{h_s}{v_s}=\frac{h}{v},
\end{eqnarray}

where $h_a$, $h_w$, $h_s$ are the ultrasonic ray path lengths in
air, in free water, and in solid phase, respectively, $h$ is the
testing piece length ($200$ $mm$), $v_a=330$ $m/s$, $v_w=1500$
$m/s$, $v_s$ is the ultrasonic propagation velocity in the
concrete solid phase, and finally, $v$ represents the effective
ultrasonic propagation velocity in the heterogeneous medium
(concrete).\\

When the testing piece does not exhibit moisture at all it can be
considered that the term $h_w/v_w$ in the equation (2) is
approximately equal to zero and the transition time $t$ of the
ultrasonic ray through the testing piece takes its maximum value
$t_{max}$. It corresponds to a minimum value of the effective
ultrasonic propagation velocity $v$ in the heterogeneous medium.
Besides, the variable $h_a$ takes its maximum value $h_p$ which
corresponds to the ultrasonic ray path length in porous space when
only the air is present. In this case, the equation (2) can be
written as

\begin{eqnarray}
t_{max}-\delta=\frac{h_p}{v_a}+\frac{h_s}{v_s}=\frac{h}{v_{min}},
\end{eqnarray}

On the other hand, if the testing piece is quite saturated with
the free water, we can consider that the term $h_a/v_a$ in the
equation (2) is approximately equal to zero, the variable $h_w$
takes its maximum value $h_p$, the variable $t$ takes its minimum
value $t_{min}$ corresponding to the maximum value of the variable
$v$. Then, the equation (2) will become to

\begin{eqnarray}
t_{min}-\delta=\frac{h_p}{v_w}+\frac{h_s}{v_s}=\frac{h}{v_{max}},
\end{eqnarray}

We have assumed in the analysis above that the quantities
$h_s$,$v_s$, and $h$ are fixed. It is also considered that

\begin{eqnarray}
h_w+h_a=h_p=constant,\\
h_p+h_s=h=constant
\end{eqnarray}

From the equations (3) and (4) we can find that

\begin{eqnarray}
h_p=(t_{max}-t_{min})/k_1,
\end{eqnarray}

where the constant $k_1=2364 \mu s/m$. By using the equation
above, we can estimate the ultrasonic ray path length in porous
space, where $t_{max}$ corresponds to the transition time of the
ultrasonic ray in the testing piece when it is saturated with free
water, and $t_{min}$ is the value of the same variable when the
testing piece is quite dried.\\

On the other hand, we can derive a relationship for the dependence
of the variable $t$ (transition time of the ultrasonic ray in the
testing piece) with the variable $h_w$ (ultrasonic ray path length
in the free water in the porous space in concrete). In fact, from
the equations (2), (3), and (5) it can be written that

\begin{eqnarray}
t_{max}-t=h_w*k_1
\end{eqnarray}

It can be observed that the correction term $\delta$ in equation
(2) cancels itself and it does not appear in equation (8).
Variables $t$ and $h_w$ in the equation (8) are well defined in
the intervals $t\in[t_{max},t_{min}]$ and $h_w\in[0,h_p]$. The
variable $t$ can be directly measured, but the variable $h_w$ can
not. Here we considered that the variable $h_w$ depends on the
quantity of free water ($m_w$) in the testing piece, but such a
dependence will not be a simple one. The way in which $h_w$
depends on $m_w$ is not a known matter for theoretical explaining,
but we suppose that such a dependence is influenced by the
morphology of pores and by the statistical distribution law of
pores in the testing piece volume, among other factors. These
aspects will be considered in further investigations. We know that
the variable $h_w$ and the variable $m_w$ are well defined in the
intervals $h_w\in[0,h_p]$ and $m_w\in[0,m_{w,max}]$. We assume the
following empirical relationship

\begin{eqnarray}
h_w=Q*m_w^{\gamma},\;\;Q>0,\;\;1>\gamma>0,
\end{eqnarray}

being $Q$ a positive dimensional constant and the exponent
$\gamma$ should be in the interval $\gamma \in(0,1)$. The
particular values which the proportional constant $Q$ and the
exponent $\gamma$ take in a particular experiment will be
determined by the factors indicated above. Introducing the
equation (9) in the equation (8) the following relationship is
obtained

\begin{eqnarray}
t_{max}-t=m_w^{\gamma}*Q*k_1,
\end{eqnarray}

where now both variables $t$ and $m_w$ can be directly measured.
It is convenient to linearizing the equation (10) by applying
natural log function to both sides of the equation,

\begin{eqnarray}
\log(t_{max}-t)=\gamma*\log(m_w)+\log(Q*k_1)
\end{eqnarray}

Applying a linear regression method to a log-log plotting of the
variables ($t_{max}-t$) and ($m_w$) the best straight line is
fitted to the experimental data point ($m_w,t_{max}-t$), and the
model parameters $\gamma$ and $Q$ can be determined. In this plot,
the point (0,0) can not be taken into account.

\section{Experimental data points and fitting function}

In order to evaluate the proposed empirical model, given by the
equation (11), we considered the experimental data points obtained
by Ohdaira and Masuzawa\cite{Masuzawa}. They measured the
transition time $t$ of an ultrasonic ray in a cylindrical testing
piece of concrete being $200$ $mm$ in length by applying the
ultrasonic transmission technique. The corresponding value of the
quantity of free water $m_w$ in the testing concrete piece was
measured by using a special device. The cotas of error for the
variables $t$ and $m_w$, and the parameter $h$, all of them
directly measured, are given. They applied the procedure to three
types of testing piece named A-1, B-1, and C-1. From the
experimental data points obtained by these authors we only
considered those whose were subjects of our analysis. Besides, the
experimental data points considered by us were organized as data
vectors because it made easier data processing when certain
software was used. Ohdaira and Masuzawa\cite{Masuzawa} measured
the transition time $t$ five times for each value of the free
water $m_w$. Statistical validation allows us to average these
five values by taking the mean value as one element (in $\mu s$)
of the data vector called $\langle time \rangle$. The vector
$\langle water \rangle$ contains the values of the variable $m_w$
in kilogram. These vectors are separately given below for each
testing piece.\\

Testing piece A-1:

\begin{eqnarray}
\nonumber water=[0.2321\; 0.2109\; 0.1969\; 0.1824\; 0.1694\;
0.1571\;0.1404 ...\\
0.1227\; 0.1060\; 0.0787\; 0.0615\; 0.0419\; 0.0256\; 0.0089\;
0.0000];\\
\nonumber time=[44.48\; 45.16\; 46.08\; 46.64\; 46.48\; 46.44\;
46.60\; 46.48 ...\\ 47.24\; 47.80\; 48.24\; 49.16\; 49.76\;51.08\;
51.32];
\end{eqnarray}

Testing piece B-1:

\begin{eqnarray}
\nonumber water=[0.2500\; 0.2165\; 0.2001\; 0.1850\; 0.1724\;
0.1614\;0.1449 ...\\ 0.1278\; 0.1093\; 0.0786\; 0.0613\; 0.0406\;
0.0240\; 0.0084\; 0.0000];\\
\nonumber time=[44.88\; 46.32\; 46.12\; 46.32\; 46.80\; 46.44\;
47.48\; 47.00 ...\\ 47.24\; 47.76\; 48.28\; 49.20\; 50.84\;
51.60\; 51.80];
\end{eqnarray}

Testing piece C-1:

\begin{eqnarray}
\nonumber water=[0.2479\; 0.2257\; 0.2021\; 0.1871\; 0.1693\;
0.1555\; 0.1436 ...\\
0.1259\; 0.1082\; 0.0896\; 0.0656\; 0.0531\; 0.0379\; 0.0241\; 0.0088\; 0.0000];\\
\nonumber time=[46.64\; 47.12\; 48.52\; 48.52\; 48.76\; 49.20\;
49.16\; 49.40 ...\\ 49.64\; 50.08\; 50.28\; 50.52\; 51.64\;
52.40\; 52.64\; 53.08];
\end{eqnarray}

We estimated the random cota of error for the vector $\langle time
\rangle$ and composed it together with the instrumental cota of
error whose value is equal to 0.2 $\mu s$\cite{Masuzawa} for
having the vector $\langle cota-time \rangle$ in $\mu s$,\\

Testing piece A-1:

\begin{eqnarray}
\nonumber
cota-time=[0.31\;0.22\;0.25\;0.34\;0.36\;0.22\;0.28\; ...\\
0.22\;0.23\;0.23\;0.28\;0.38\;0.30\;0.22\;0.24];
\end{eqnarray}

Testing piece B-1:

\begin{eqnarray}
\nonumber
cota-time=[0.21\;0.22\;0.21\;0.21\;0.23\;0.23\;0.24\; ...\\
0.26\;0.22\;0.26\;0.26\;0.22\;0.28\;0.29\;0.22];
\end{eqnarray}

Testing piece C-1:

\begin{eqnarray}
\nonumber
cota-time=[0.20\;0.22\;0.24\;0.21\;0.24\;0.22\;0.23\; ...\\
0.22\;0.23\;0.27\;0.23\;0.24\;0.34\;0.32\;0.54\;0.28];
\end{eqnarray}

The vector $\langle time \rangle$ was corrected by the constant
value $\delta=0.6$ $\mu s$ (equation (1)) for obtaining the vector
$\langle timec \rangle$ in $\mu s$,\\

Testing piece A-1:

\begin{eqnarray}
\nonumber
timec=[43.88\;44.56\;45.48\;46.04\;45.88\;45.84\;46.00\;...\\
45.88\;46.64\;47.20\;47.64\;48.56\;49.16\;50.48\;50.72];
\end{eqnarray}

Testing piece B-1:

\begin{eqnarray}
\nonumber timec=[44.28\;45.72\;45.52\;45.72\;46.20\;45.84\;46.88\; ...\\
46.40\;46.64\;47.16\;47.68\;48.60\;50.24\;51.00\;51.20];
\end{eqnarray}

Testing piece C-1:

\begin{eqnarray}
\nonumber timec=[46.04\;46.52\;47.92\;47.92\;48.16\;48.60\;48.56\; ...\\
48.80\;49.04\;49.48\;49.68\;49.92\;51.04\;51.80\;52.04\;52.48];
\end{eqnarray}

By using the vector $\langle timec \rangle$ the vector $\langle
velocity \rangle$ was constructed by applying the relationship

\begin{eqnarray}
v=h/(t-\delta)
\end{eqnarray}

where $h=200$ $mm$ is the length of each testing piece and
$t-\delta$ are the elements of the vector $\langle timec \rangle$.
The elements of the vector $\langle velocity \rangle$ are given in
$m/s$,\\

Testing piece A-1:

\begin{eqnarray}
\nonumber
velocity=[4557.9\;4488.3\;4397.5\;4344.0\;4359.2\;4363.0\;4347.8\;...\\
4359.2\;4288.2\;4237.3\;4198.2\;4118.6\;4068.3\;3962.0\;3943.2];
\end{eqnarray}

Testing piece B-1:

\begin{eqnarray}
\nonumber
velocity=[4516.7\;4374.5\;4393.7\;4374.5\;4329.0\;4363.0\;4266.2\;...\\
4310.3\;4288.2\;4240.9\;4194.6\;4115.2\;3980.9\;3921.6\;3906.3];
\end{eqnarray}

Testing piece C-1:

\begin{eqnarray}
\nonumber
velocity=[4344.0\;4299.2\;4173.6\;4173.6\;4152.8\;4115.2\;4118.6\; ...\\
4098.4\;4078.3\;4042.0\;4025.8\;4006.4\;3918.5\;3861.0\;3843.2\;3811.0];
\end{eqnarray}

Finally, the cotas of error for the transition time and for the
length of the testing piece ($\delta h=0.0005$ $m$) were
propagated through the formula (24) written in the form of finite
increment,

\begin{eqnarray}
\delta v=\frac{h}{(t-\delta)^2}\;\delta t+
\frac{1}{(t-\delta)}\;\delta h
\end{eqnarray}

The vector $\langle \delta velocity \rangle$ in $m/s$ represents
the cota of error vector of the vector $\langle velocity \rangle$
propagated by the equation above. For each type of testing piece
we obtained,\\

Testing piece A-1:

\begin{eqnarray}
\nonumber \delta
velocity=[43.31\;33.65\;35.15\;42.66\;45.21\;32.10\;36.93 ...\\
31.36\;32.01\;31.06\;35.04\;42.21\;34.94\;27.52\;28.65];
\end{eqnarray}

Testing piece B-1:

\begin{eqnarray}
\nonumber \delta
velocity=[32.30\;31.55\;30.86\;30.64\;32.19\;32.94\;32.66 ...\\
35.00\;31.20\;34.33\;33.16\;28.84\;32.03\;32.09\;26.48];
\end{eqnarray}

Testing piece C-1:

\begin{eqnarray}
\nonumber \delta
velocity=[30.10\;30.66\;31.48\;28.37\;31.08\;28.84\;29.93 ...\\
28.65\;29.45\;31.78\;28.96\;29.41\;35.67\;33.69\;49.35\;29.86];
\end{eqnarray}

By considering the equation (11) the log-log plotting was applied
to each of the three types of testing pieces. Figures 1, 2 and 3
show the results. The best straight line fitting the experimental
data points produced the following values for the model parameters
$\gamma$ and $Q$,

\begin{eqnarray}
\gamma_{A-1}=0.8707,\;\;\gamma_{B-1}=0.9205,\;\;\gamma_{C-1}=0.7923,\\
Q_{A-1}=0.0111,\;\;Q_{B-1}=0.0124,\;\;Q_{C-1}=0.0080.
\end{eqnarray}

Figure 4 depicts the plotting of the equation (9) using the values
of the model parameters indicated above.\\

On the other hand, by applying the equation (7), the parameter
$h_p$ in meter was estimated

\begin{eqnarray}
h_{p,A-1}=0.0029,\;\;h_{p,B-1}=0.0029,\;\;h_{p,C-1}=0.0027
\end{eqnarray}

From the equation (2) and (10), the effective ultrasonic pulse
propagation velocity $v$ as a function of $m_w$ is given by

\begin{eqnarray}
v=\frac{h/k_2}{(1-(Qk_1/k_2)m_w^{\gamma})},\;\;k_2=t_{max}-\delta
\end{eqnarray}

This equation is the fitting function for the experimental data
points ($m_w$,$v$). Using the numerical values of the parameters
$h$, $k_2$, $Q$, $k_1$, and $\gamma$, the equation (35) for each
of the testing pieces are respectively,

\begin{eqnarray}
v_{A-1}=\frac{3943.22}{(1-(0.52524)m_w^{0.8707})},\\
v_{B-1}=\frac{3906.25}{(1-(0.58125)m_w^{0.9205})},\\
v_{C-1}=\frac{3810.98}{(1-(0.365854)m_w^{0.7923})}
\end{eqnarray}

Figures 5, 6, and 7 separately depict the experimental data points
($m_w$,$v$) and the fitting function given in each case by the
equations (36), (37), and (38).

\section{Discussion of the results}

The influence of moisture in the testing piece of concrete on the
effective ultrasonic pulse propagation velocity is highly
significant. Experimental results confirm this fact. Though the
ultrasonic transmission technique for measuring the effective
ultrasonic pulse propagation velocity does not exhibit high
accuracy, it does allow to determine the level of such an
influence and the behavior of the effective ultrasonic pulse
propagation velocity with water content can be well defined. The
functional form by which such an influence is determined depends,
among other factors, on the morphology of pores and on the
statistical distribution law of pores in the testing piece volume.
However, we have considered an empirical relationship between the
variable $h_w$ and the variable $m_w$ that indirectly very well
matches to experimental data points through the relationship (35)
between the effective ultrasonic pulse propagation velocity $v$
and the quantity of free water $m_w$ in the testing piece. The
finding of this fitting function has been the aim of this work.
Figure 4 depicts the empirical relationship between $h_w$ and
$m_w$ for each type of testing pieces. It is interesting to
observe that the values of the parameters $\gamma$ and $Q$
increases with the increase in cement content in the testing piece
in the considered range. It is just a conjecture. This aspect will
be considered in further investigations.\\

We observe that the relationship between $v$ and $m_w$ is not a
linearly proportional function. Figures 5, 6 and 7 for testing
pieces A-1, B-1, and C-1, show a suitable fitting of the
experimental data points by the equation (35).\\

It is important to check whether the proposed model correctly
describes the behavior of the experimental data ($v=f(m_w)$). The
equation (35) can not be linearized, but if $m_w=0$, then the
velocity coincides with the minimum velocity $v_0$ (velocity in
dry state) and, consequently, equation (35) can be written in the
form

\begin{eqnarray}
y=\frac{a}{1-b x^c},
\end{eqnarray}

where $y=v$ is the effective ultrasonic pulse propagation
velocity, $a=v_0$ represents the minimum velocity (in dry state),
$x=m_w$ is the free water content, $b=Q(k1/k2)$ and $c=\gamma$ are
parameters related with the parameters indicated in the
equation(35). When $x=0$, from equation (39) is obtained that
$y=a$.\\

The results of the statistical analysis are presented in table
no.1, for each one of the samples $A_1$, $B_1$, and $C_1$.\\

The model very well fits the experimental data, and the values of
parameter $c$ do not significatively differ from those of $\gamma$
obtained in the equation (32). In all the cases they are within
the deviation range proposed by the model. One can conclude that
the model given in equations (36, 37, 38), appropriately describes
the experimental data set.\\

Next, we analyze under which conditions the behavior of the
variable $v$ with the variable $m_w$ can be considered a linearly
proportional relationship. From the equation (35) we can obtain
the following relationship by applying the Newtonian binomial
formula. The quadratic approximation is

\begin{eqnarray}
v^{(2)}\cong(h/k_2)*(1+(Qk_1/k_2)m_w^{\gamma}+(Qk_1/k_2)^2*m_w^{2\gamma}),\;\;
(Qk_1/k_2)m_w^{\gamma}<<1
\end{eqnarray}

The linear approximation of the equation above is

\begin{eqnarray}
v^{(1)}\cong(h/k_2)*(1+(Qk_1/k_2)m_w^{\gamma}),\;\;
(Qk_1/k_2)m_w^{\gamma}<<1
\end{eqnarray}

and the contribution to the effective ultrasonic propagation
velocity of the quadratic term is,

\begin{eqnarray}
v^{(2)}-v^{(1)}\cong(h/k_2)*(Qk_1/k_2)*m_w^{\gamma}
\end{eqnarray}

The vector $\langle errorv \rangle$ in $m/s$ contains the
contributions of the quadratic term in equation (42). For each
testing piece we can write,\\

Testing piece A-1:

\begin{eqnarray}
\nonumber \langle errorv
\rangle=[82.28\;69.64\;61.79\;54.08\;47.55\;41.70\;34.29
...\\
27.12\;21.02\;12.51\;8.15\;4.18\;1.77\;0.28\;0.00];
\end{eqnarray}

Testing piece B-1:

\begin{eqnarray}
\nonumber \langle errorv
\rangle=[99.72\;76.52\;66.19\;57.29\;50.31\;44.56\;36.53 ...\\
28.99\;21.74\;11.85\;7.50\;3.51\;1.33\;0.19\;0.00];
\end{eqnarray}

Testing piece C-1:

\begin{eqnarray}
\nonumber \langle errorv
\rangle=[53.77\;46.34\;38.90\;34.43\;29.38\;25.68\;22.63
...\\
18.38\;14.45\;10.72\;6.54\;4.68\;2.74\;1.34\;0.27\;0.00];
\end{eqnarray}

We can compare the vectors $\langle \delta velocity \rangle$ and
$\langle errorv \rangle$ for deciding when the equation (41) is
suitable to apply. When $m_w\geq (0.16)$ in kilogram it can be
observed that the contribution of the quadratic term (equation 42)
is greater than the propagated cota of error for the experimental
value of the effective ultrasonic propagation velocity $v$ given
by the equation (28). The equation (41), considering that the
exponent $\gamma\cong 1$, approximately well describes the linear
behavior considered in \cite{Masuzawa}, but this approach implies
a greater error in velocity when $m_w$ is approximately greater
than $0.16$ kg.\\

On the other hand, the equation (35) very well quantitatively
describes the effect of moisture in concrete on $v$. In fact, the
formula (35) can be conveniently written as

\begin{eqnarray}
\frac{\delta v}{v}=\frac{k_1*h}{(t_{max}-h_w*k_1)^2}(100*h_p/h)
\end{eqnarray}

where $100*h_p/h$ represents the ratio in percent between the
ultrasonic ray path length in porous space and the testing piece
length. Applying this equation to experimental data points for
each testing piece of concrete it may be found that such a small
value of the ratio $100*h_p/h$ as $\cong 1.45$ percent can yield a
highly significative value of the fraction $\delta v/v$,
approximately $14-16$ percent. This result is well confirmed by
the experimental data points ($v,m_w$).\\

\section{Conclusions}

The result of this analysis confirms that the empirical fitting
function very well quantitatively describes the behavior of
effective ultrasonic pulse propagation velocity with the free
water content in concrete testing piece and that this empirical
model well explains under which condition the linear approximation
can be applied. It also allows to estimate the propagation of
error for the indirectly-measured involved variables.\\

\section*{Acknowledgement}

We acknowledge Pedro A. S\'anchez Fern\'andez from the Department
of Foreign Languages for the revision of the English version, and
Yoelsy Leiva for the edition. We also thank the Ministry of Higher
Education of Cuba for partially financial support of the research.
Finally, we express our thankfulness to E. Ohdaira and N. Masuzawa
for publishing their experimental data points.

\newpage
\section{Tables}
\newpage
\ Table no.1: Test Piece A-1
\newline

\begin{tabular}{|c|c|c|c|c|}
 \hline R  & R.squared & R.squared adj & Typical error of the estimate & Durbin-Watson
  \\ \hline
 .955 & .912 & .905 & .26315 & 1.187 \\ \hline
\end{tabular}

\begin{center}
\begin{tabular}{|c|c|c|c|c|c|c|c|}
  \hline  & Non  &  & Standardized   & t & Sig. &Confidence   &   \\
      & standardized  & &Coefficients  &  &  & Interval 95 \% &  \\\hline
       & Coefficients  & &  &  &  &  &  \\\hline
   &B & Typ. Error & Beta & &  & Lower  bound & Upper bound \\ \hline
  Ln b´ & -8.944 &.197 &  & -45.314 & .000 & -9.374 &-8.514 \\ \hline
  C & .870 & .078 & .955 & 11.177 & .000 & .700 & 1.040 \\ \hline
\end{tabular}
\end{center}

\newpage
\ Table no.1: Test Piece B-1
\newline

\begin{tabular}{|c|c|c|c|c|}
 \hline R  & R.squared & R.squared adj & Typical error of the estimate & Durbin-Watson
  \\ \hline
 .948 & .898 & .89 & .31323 & 1.667 \\ \hline
\end{tabular}

\begin{center}
\begin{tabular}{|c|c|c|c|c|c|c|c|}
  \hline  & Non  &  & Standardized   & t & Sig. &Confidence   &   \\
      & standardized  & &Coefficients  &  &  & Interval 95 \% &  \\\hline
       & Coefficients  & &  &  &  &  &  \\\hline
   &B & Typ. Error & Beta & &  & Lower  bound & Upper bound \\ \hline
  Ln b´ & -8.828 &.227 &  & -38.867 & .000 & -9.323 &-8.333 \\ \hline
  C & .921 & .089 & .948 & 10.294 & .000 & .726 & 1.116 \\ \hline
\end{tabular}
\end{center}

\newpage
\ Table no.1: Test Piece C-1
\newline

\begin{tabular}{|c|c|c|c|c|}
 \hline R  & R.squared & R.squared adj & Typical error of the estimate & Durbin-Watson
  \\ \hline
 .977 & .955 & .951 & .16848 & 1.265 \\ \hline
\end{tabular}

\begin{center}
\begin{tabular}{|c|c|c|c|c|c|c|c|}
  \hline  & Non  &  & Standardized   & t & Sig. &Confidence   &   \\
      & standardized  & &Coefficients  &  &  & Interval 95 \% &  \\\hline
       & Coefficients  & &  &  &  &  &  \\\hline
   &B & Typ. Error & Beta & &  & Lower  bound & Upper bound \\ \hline
  Ln b´ & -9.271 &.122 &  & -75.767 & .000 & -9.535 &-9.006 \\ \hline
  C & .793 & .048 & .977 & 16.557 & .000 & .689 & .896 \\ \hline
\end{tabular}
\end{center}






\newpage
\section{Figures}

\newpage
\begin{figure}[tbh!]
\centerline{\psfig{figure=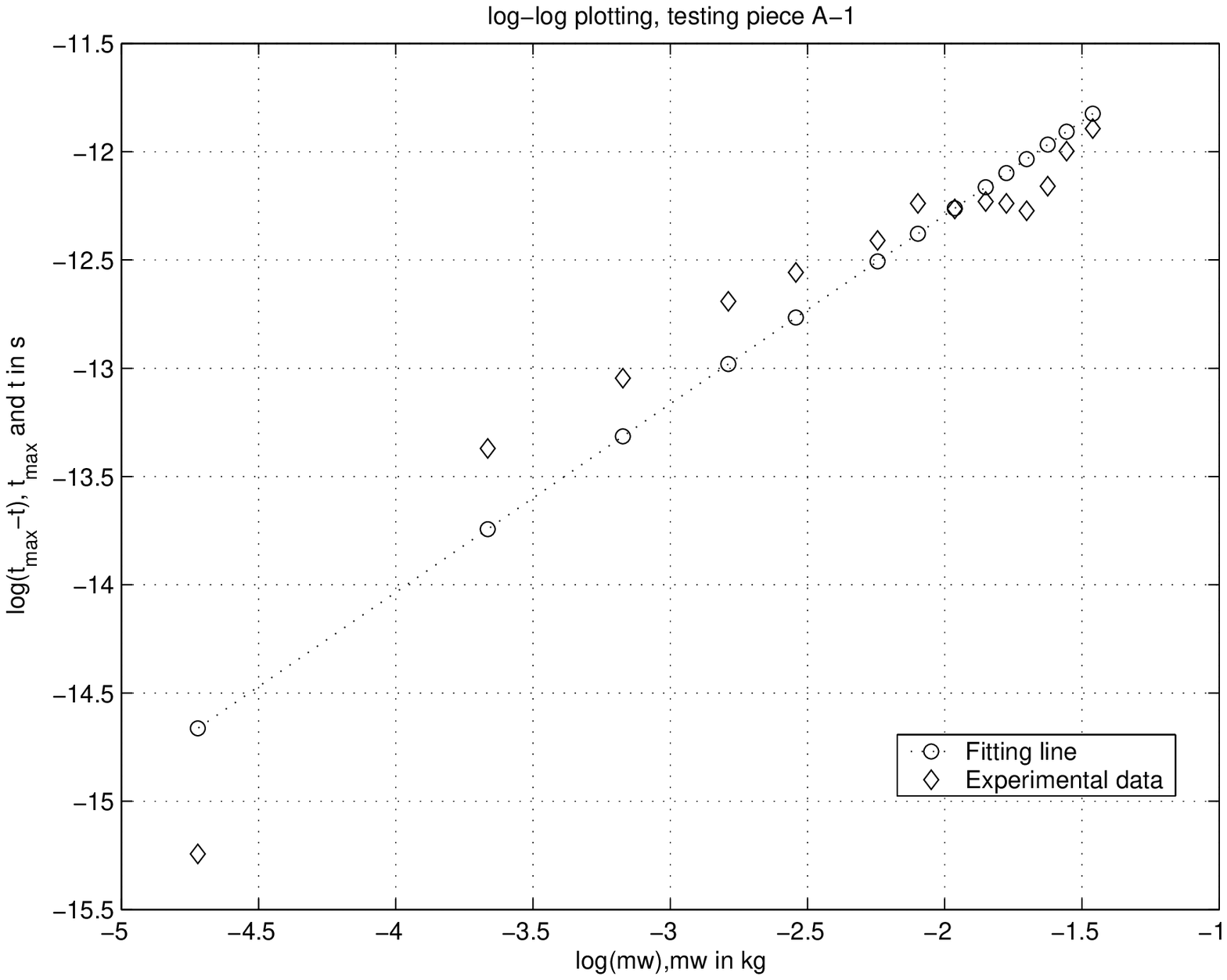,width=0.5\textwidth,angle=0}}
\bigskip
\caption{Linear regression, testing piece A-1} \label{fig1}
\end{figure}

\begin{figure}[tbh!]
\centerline{\psfig{figure=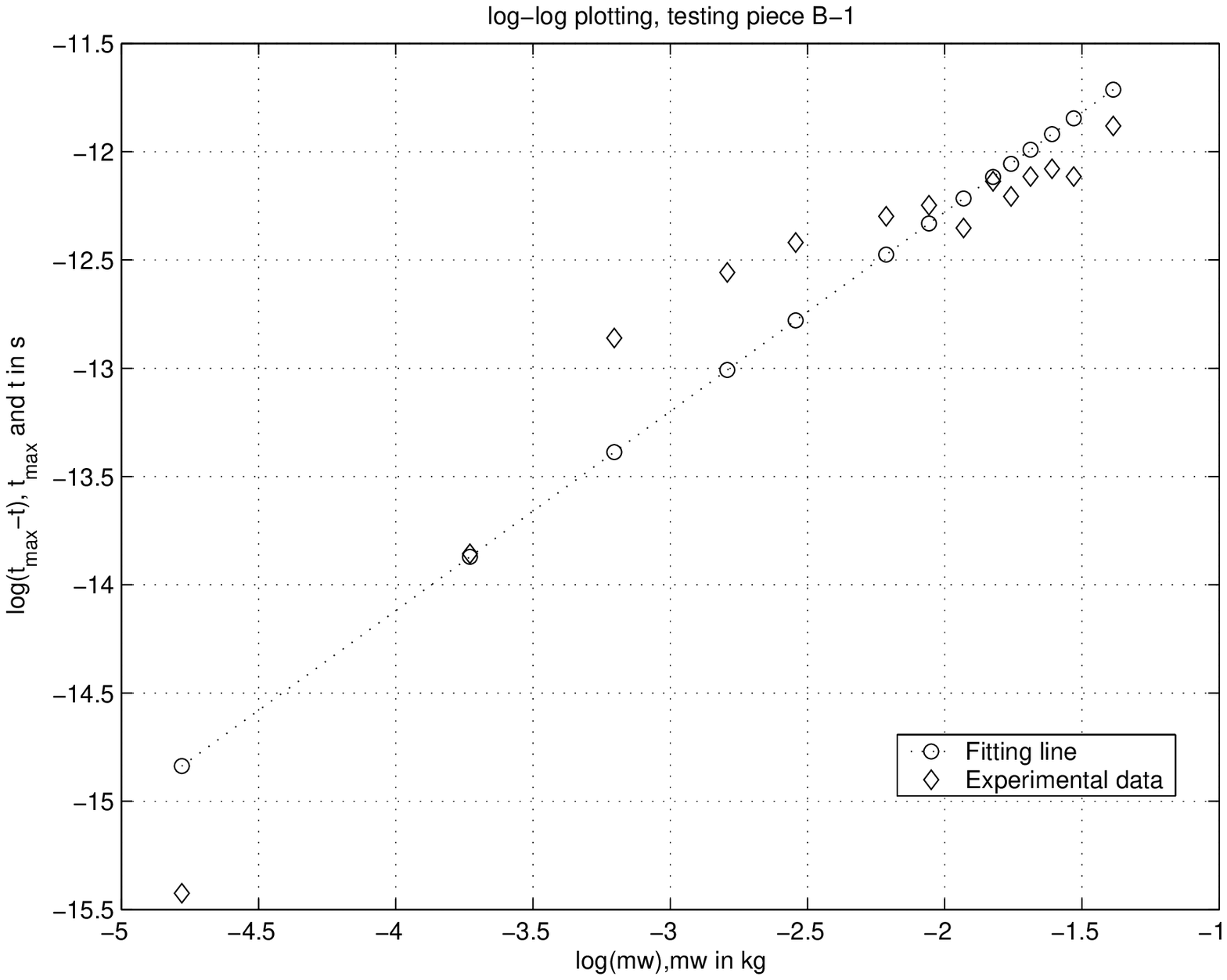,width=0.5\textwidth,angle=0}}
\bigskip
\caption{Linear regression, testing piece B-1} \label{fig2}
\end{figure}

\begin{figure}[tbh!]
\centerline{\psfig{figure=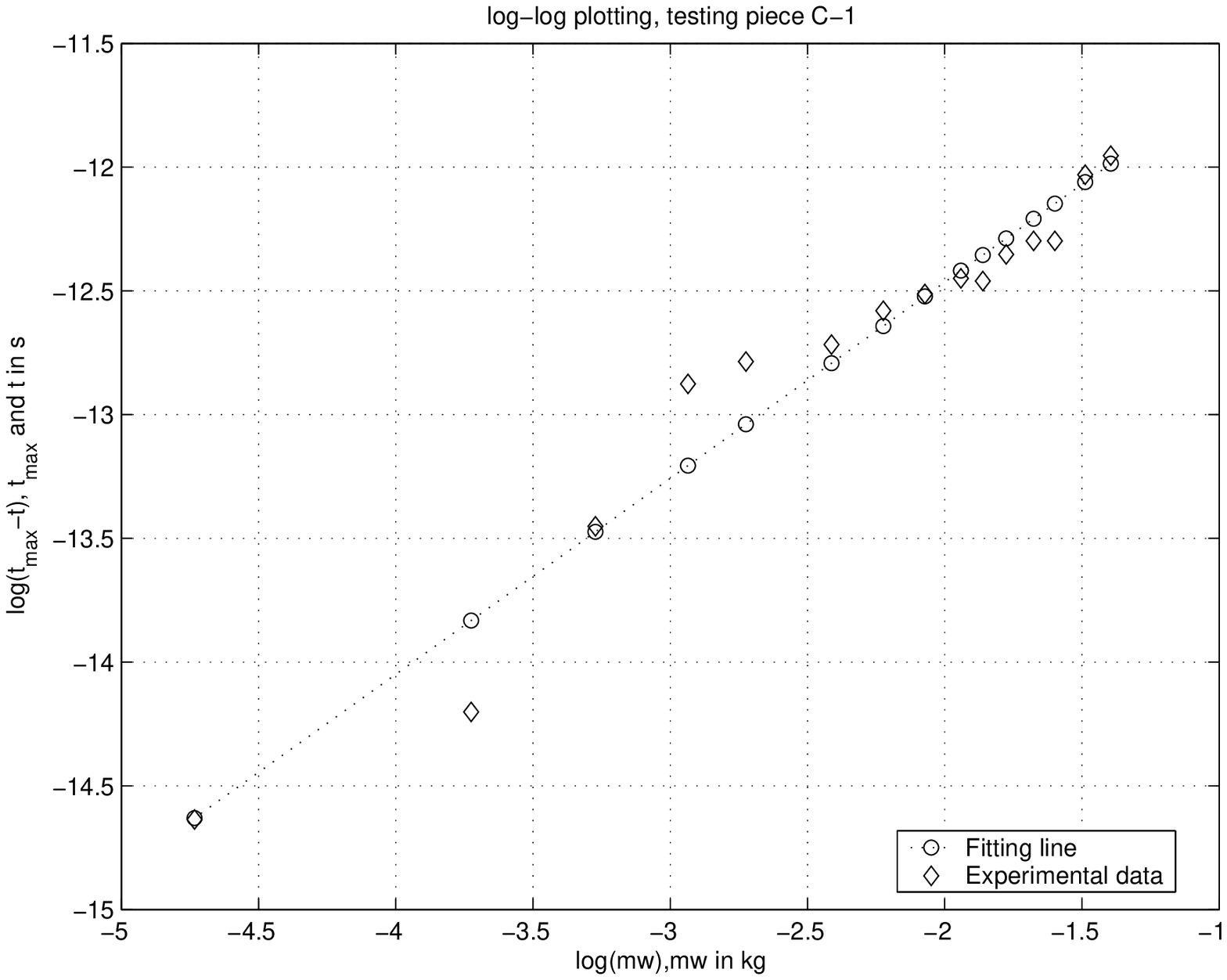,width=0.5\textwidth,angle=0}}
\bigskip
\caption{Linear regression, testing piece C-1} \label{fig3}
\end{figure}

\begin{figure}[tbh!]
\centerline{\psfig{figure=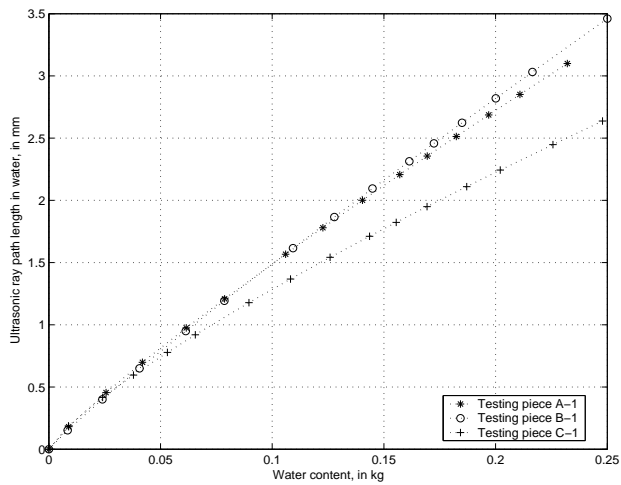,width=0.5\textwidth,angle=0}}
\bigskip
\caption{Curves of the empirical model: ultrasonic ray path length
in the free water in the porous space in concrete versus free
water content} \label{fig4}
\end{figure}

\begin{figure}[tbh!]
\centerline{\psfig{figure=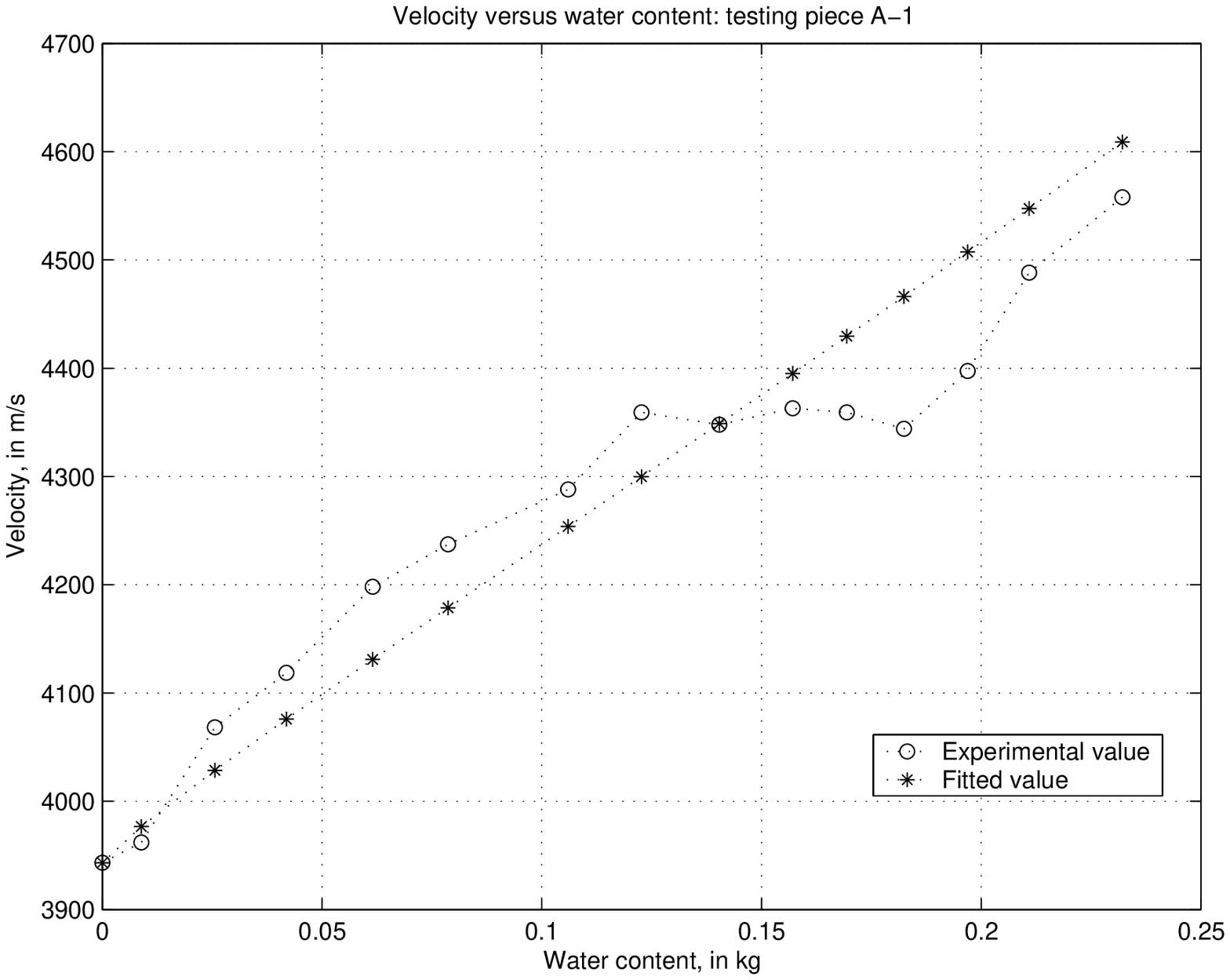,width=0.5\textwidth,angle=0}}
\bigskip
\caption{Fitting function, testing piece A-1} \label{fig5}
\end{figure}

\begin{figure}[tbh!]
\centerline{\psfig{figure=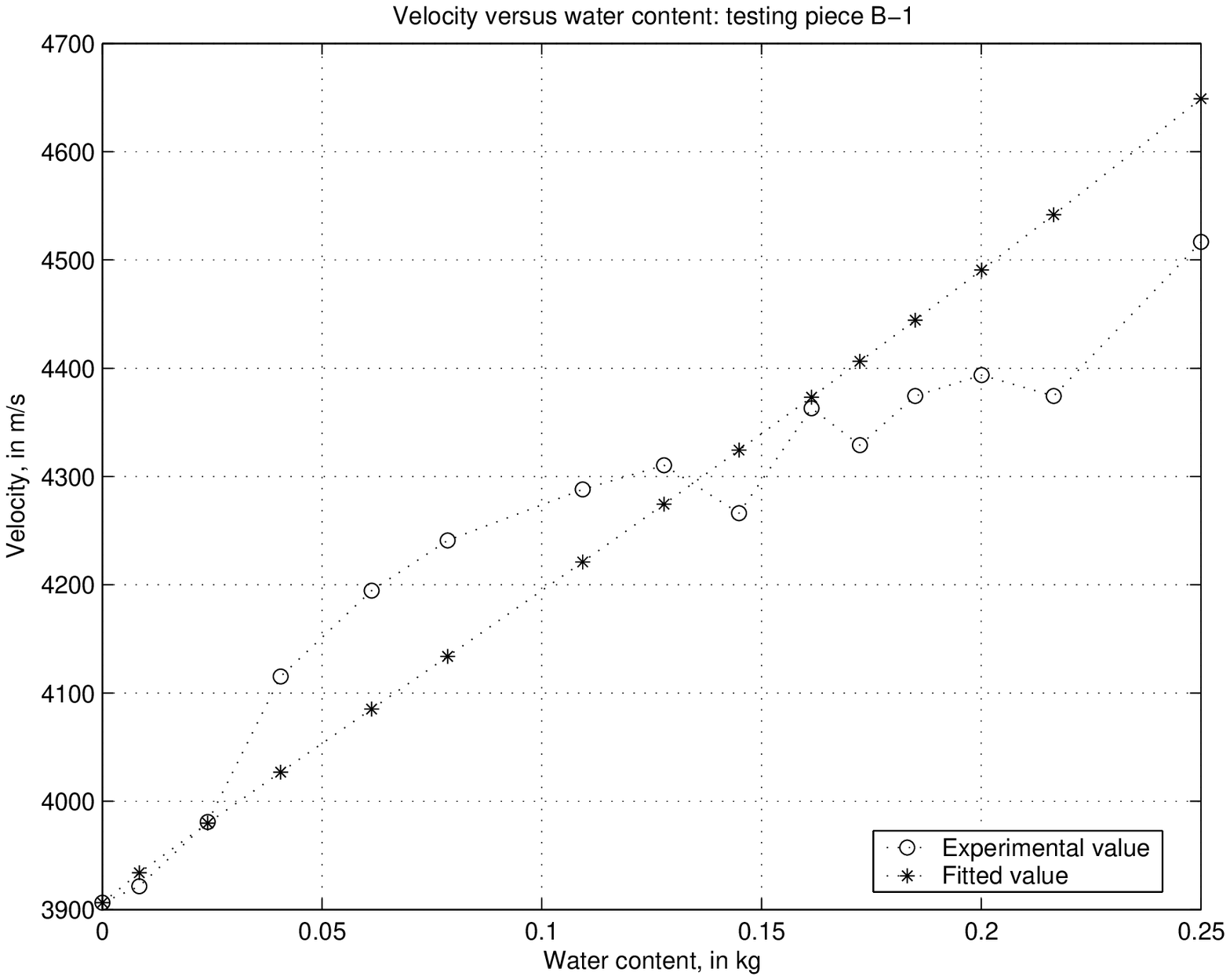,width=0.5\textwidth,angle=0}}
\bigskip
\caption{Fitting function, testing piece B-1} \label{fig6}
\end{figure}

\begin{figure}[tbh!]
\centerline{\psfig{figure=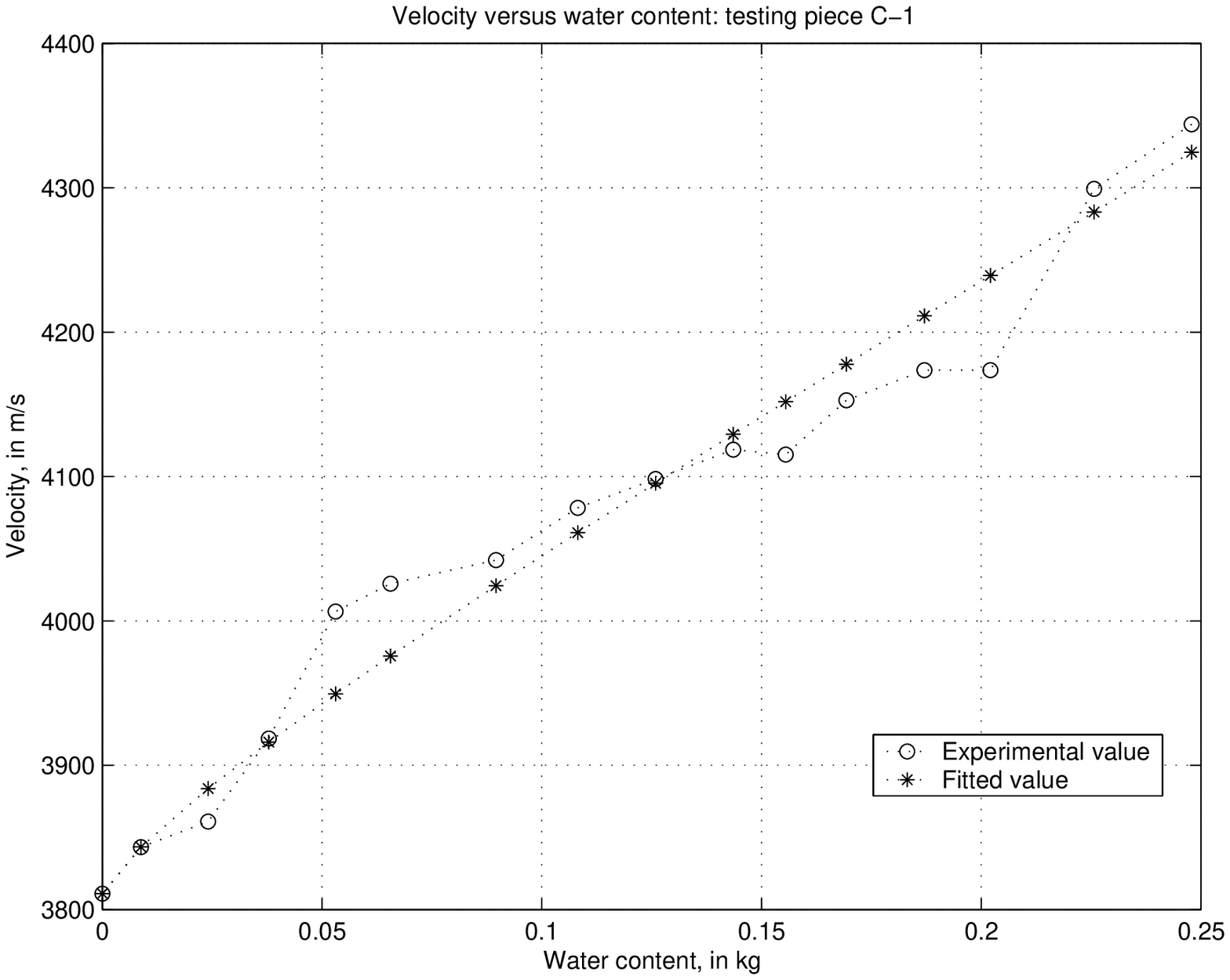,width=0.5\textwidth,angle=0}}
\bigskip
\caption{Fitting function, testing piece C-1} \label{fig7}
\end{figure}

\end{document}